\documentclass[preprint]{elsarticle}
\usepackage{graphicx,url,hyperref,microtype,subcaption,graphbox,amsmath}
\usepackage{latexsym,courier,upgreek,xcolor,rotating,ulem}
\usepackage[onehalfspacing]{setspace}
\hypersetup{colorlinks,citecolor=blue,linkcolor=red,urlcolor=green}
\usepackage[displaymath,mathlines]{lineno}
\usepackage[margin=2cm]{geometry}
\modulolinenumbers[5]
\biboptions{numbers,sort&compress}

\begin{document}

\begin{frontmatter}
  
  \title{Record statistics of emitted energies $-$ prediction of an upcoming failure}
  
  \author[add1]{Subhadeep Roy}
  \ead{subhadeep.r@hyderabad.bits-pilani.ac.in}
  \address[add1]{Dept. of Physics, Birla Institute of Technology \& Science Pilani, Hyderabad Campus, Secunderabad, Telangana 500078.}

  \begin{abstract}
The article reports a numerical investigation of the breakdown of a disordered system considering the effect of local stress concentration under the action of an external tensile force. The statistics of the record-breaking magnitudes of emitted energies during the failure process as well as the waiting time to achieve those record events show rich behavior. The latter includes information about the acceleration and subsequent catastrophic failure through its non-monotonic behavior. The maximum waiting time is also correlated with the maximum change in elastic energy as the model evolves, a different way of predicting an upcoming failure, which is consistent with our hypothesis as well. At a moderate disorder, such a prediction can be done with higher accuracy while at a low disorder, due to the abrupt nature of the failure process our hypothesis does not hold well.  
  \end{abstract}

  \date{\today}
  
  
\end{frontmatter}


\makeatletter
\def\ps@pprintTitle{%
 \let\@oddhead\@empty
 \let\@evenhead\@empty
 \def\@oddfoot{}%
 \let\@evenfoot\@oddfoot}
\makeatother


\section{Introduction}
The fracture process in a disordered system has been explored widely from the point of view of statistical mechanics due to its practical application as well as fundamental theoretical problems \cite{hr90}. One interesting aspect in this regard is the fact that heterogeneous materials do not break at once when applied to external mechanical stress. 

The micro and mesoscale heterogeneity of materials has the consequence that their fracture process evolves through local breaking events, known as intermittent bursts, that produce crackling noise which can be captured through an acoustic emission (AE) experiment \cite{b13}. The crackling noises, when translated into bursts and subsequently emitted energies, can be exploited to predict an upcoming catastrophic event \cite{n14v15}. Such acoustic signals (also known as precursors) become more and more populated closer to the global failure and approach the catastrophic event in an accelerating manner indicated by the increasing rate of deformation \cite{ppvac94}. During the initiation and propagation of cracks through a disordered media, complexity arises due to material disorders such as the distribution of flaws, dislocations, and micro-cracks present. As the external stress is increased, micro-cracks generate randomly and the stress ﬁelds of such micro-cracks interact with each other making this process correlated. 

The intensity and density of the precursory activity and its accuracy for an upcoming failure process depend on the strength of the disorder. In one limiting case, where the disorder strength is close to zero, the catastrophic failure occurs suddenly without any prior warning. In this limit, no precursors are observed \cite{zrsv97}. On the other hand, when the disorder strength is sufficiently high, a gradual accumulation of damage is observed showing bursts and avalanches with a rate growing with time \cite{rccv13}.

Examples of such avalanches are observed on different scales, starting from laboratory scales like fracturing of wood \cite{ggbc97}, cellular glass \cite{mvfs98}, concrete \cite{ppvac94}, etc., to a much larger scale of seismic events where this phenomenon is reminiscent of the Gutenberg-Richter law \cite{gr94}. Apart from failure, there are many other statistical models that incorporate avalanches such as magnetic materials showing the Barkhausen effect \cite{zcds98} and flux lines in superconductors \cite{fwnl95}, friction \cite{cl94}, Heines jumps during the flow of fluid through a porous media \cite{tkr87}, and so on. Therefore, understanding the physical mechanisms of precursor activities and avalanche dynamics goes well beyond the study of fracture and breakdown. 

One robust behavior that is observed during a failure process is how the emitted energies, captured during an acoustic emission (AE) experiment, are distributed statistically. In most cases, the distribution of emitted energy shows a scale-free behavior with a non-universal exponent that depends on the exact nature of the material. For example, a value 1.41 for human teeth \cite{wcjs21}, 1.25 in paper \cite{sta03}, 1.3 in synthetic plaster \cite{ppvac94}, 1.51 in wood and 2.0 in fiber-glass \cite{ggbc97}, 1.9 in fiberglass \cite{gcgzs02} and 1.5 in cellular glass \cite{mvfs98} etc.  

Apart from experiments, a considerable amount of effort has been given to large-scale simulations of lattice models \cite{hr90} for a better understanding of an evolving disordered system and the subsequent failure associated with the motion. Such a lattice model can be an electrical analog like a random resistor network \cite{hf77}, a mechanical analog like a random spring network \cite{mbr17}, or other more complex systems like continuum models for damage mechanics \cite{bmp22}. The lattice models usually evolve by solving a number of coupled equations (depending on the complexity of the model). A long-range interaction is introduced in these models in terms of disorder incorporated through random threshold values or bond dilution. A scaling concept \cite{hh94} can also be extracted from the lattice models through their description of topological properties during the failure process.   

A relatively simpler lattice model like the fiber bundle model (FBM) shows a scale-free avalanche size distribution in the mean-field limit with a universal exponent of 5/2 \cite{hh92}. Also, a relatively lower exponent (= 3/2) is found in the mean-field limit when the model is close to the failure point \cite{phh05}. Though, as the FBM deviates from the mean-field limit and the local stress concentration starts playing a major role in the failure process, the distribution becomes exponential instead of scale-free \cite{khh97}. A recent study on emitted energies during the failure process of an FBM shows that the distribution of energies is scale-free irrespective of what the distribution of avalanche is. The energy distribution shows a slope (= 5/2) similar to the avalanche distribution in a mean-field limit and a higher slope (= 3.5) in presence of local stress concentration \cite{rb21}. Moreover, with such local stress concentration, the model shows different modes \cite{rbr17,rhr22,srh21,srh20} of failure as well as fluctuation in failure time \cite{rbr19}.

In this paper, I will lead with a previous work where the record statistics of the avalanche sizes are explored in the context of the mean-field FBM \cite{kpk20,dk14} and later in presence of heterogeneous stress field \cite{kdpk22}. Apart from this, record statistics have been explored during compressive failure of porous materials \cite{prlkm16} as well as in the context of mining collapse \cite{jlms17}. In these articles, a warning of a catastrophic event is reported which is not complete without correlating the failure point with an observable system parameter. Moreover, I have used emitted energies as the recorded event, and not the avalanches, since it's rather realistic as well as convenient to capture energies through an Acoustic Emission (AE) experiment. A couple of recent papers \cite{rb21,bbr22} have discussed the correlation between avalanche and energy emitted during the failure of a fiber bundle model and how considering energy is more advantageous from a statistical point of view especially when the stress field is heterogeneous in space. These modifications make the present work much more relevant to real-life scenarios making it available for future experimental validation.


\section{Description of Fiber Bundle Model}

A conventional fiber bundle model \cite{Pierce,hansenBook} consists of a number of parallel Hookean fibers attached between two bars (in case of 1d) or plate (in case of 2d) which are pulled apart. If there are $L$ fibers and the externally applied stress is $F$ then the stress a single fiber experience is $\sigma=F/L$. The disorder is introduced in the model in terms of fluctuation in local strength values (say $h$) from fiber to fiber. In the present work, I have considered the following distribution 
 
\begin{equation}\label{eq2}
p(h) \sim \begin{cases}
    h^{-1},  & (10^{-\beta} \le h \le 10^{\beta}) \\
    0.  & ({\rm otherwise})
  \end{cases}
\end{equation}
Here $\beta$ is related to the span as well as the disorder strength - the higher the $\beta$ value higher the disorder strength. Such a power-law distribution has been observed earlier in local material strength values \cite{ft28}. If the local stress of a certain fiber exceeds its threshold value, that fiber breaks irreversible and the stress carried by the broken fiber is redistributed.

We have adopted here a local load-sharing scheme \cite{Phoenix,Smith} for the redistribution where the extra stress due to the rupture is redistributed among the nearest neighbors only. Since the model we consider is 1d, there are only two nearest neighbors (the redistribution scheme is elaborated in the `Method' section towards the end of this article). The other extreme is the mean-field limit where the stress of the broken fiber is redistributed equally among all surviving fibers. Apart from these two extreme limits, an intermediate stress redistribution scheme \cite{brr15,hmkh02} has also been adopted that allows a scale-free stress relaxation making the amount of stress to be redistributed a function of distance from the broken fiber. Further rupture can happen at this point without increasing the applied stress as a result of the stress redistribution, which makes the local stress values elevate. This starts an avalanche which might lead to a global failure or to the next steady state with a fraction of fibers broken and the other fraction intact. In the latter situation, we increase the applied stress just enough to break the next weakest fiber only, starting a new avalanche. Such a process goes on until all fibers are broken. The last stress at which the model attains global failure is known as the critical stress of the system.


\section{Numerical Results} 
Numerically, I have studied a fiber bundle model with system size varying between $10^3$ and $10^7$. The disorder is varied by tuning the parameter $\beta$ from 0.5 to 4.0. A low $\beta$ suggests a very low disorder where the failure process is observed to be brittle-like abrupt \cite{rr15,r17,r21}. On the other hand, for a high disorder, the bundle breaks in a number of avalanches \cite{rbr17} with energies emitted in every avalanche \cite{rb21}. This letter deals with the `record events' associated with the energies emitted. We will discuss a unique way to connect the statistics of the record energy bursts with the critical value of the applied stress at which the bundle hits the global failure. 
\begin{figure}[ht]
\centering
\includegraphics[width=7.5cm, keepaspectratio]{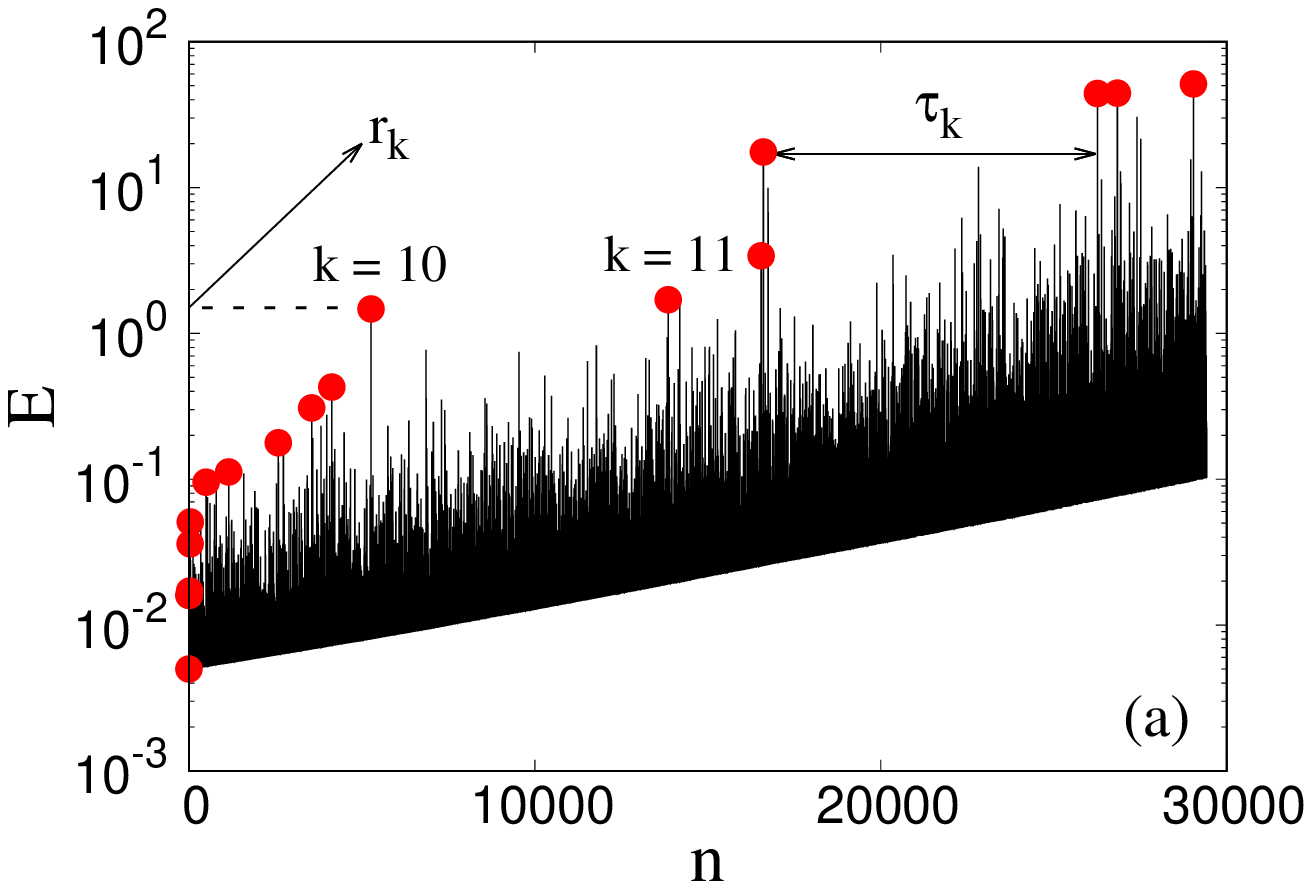} \ \ \ \ \includegraphics[width=7.2cm, keepaspectratio]{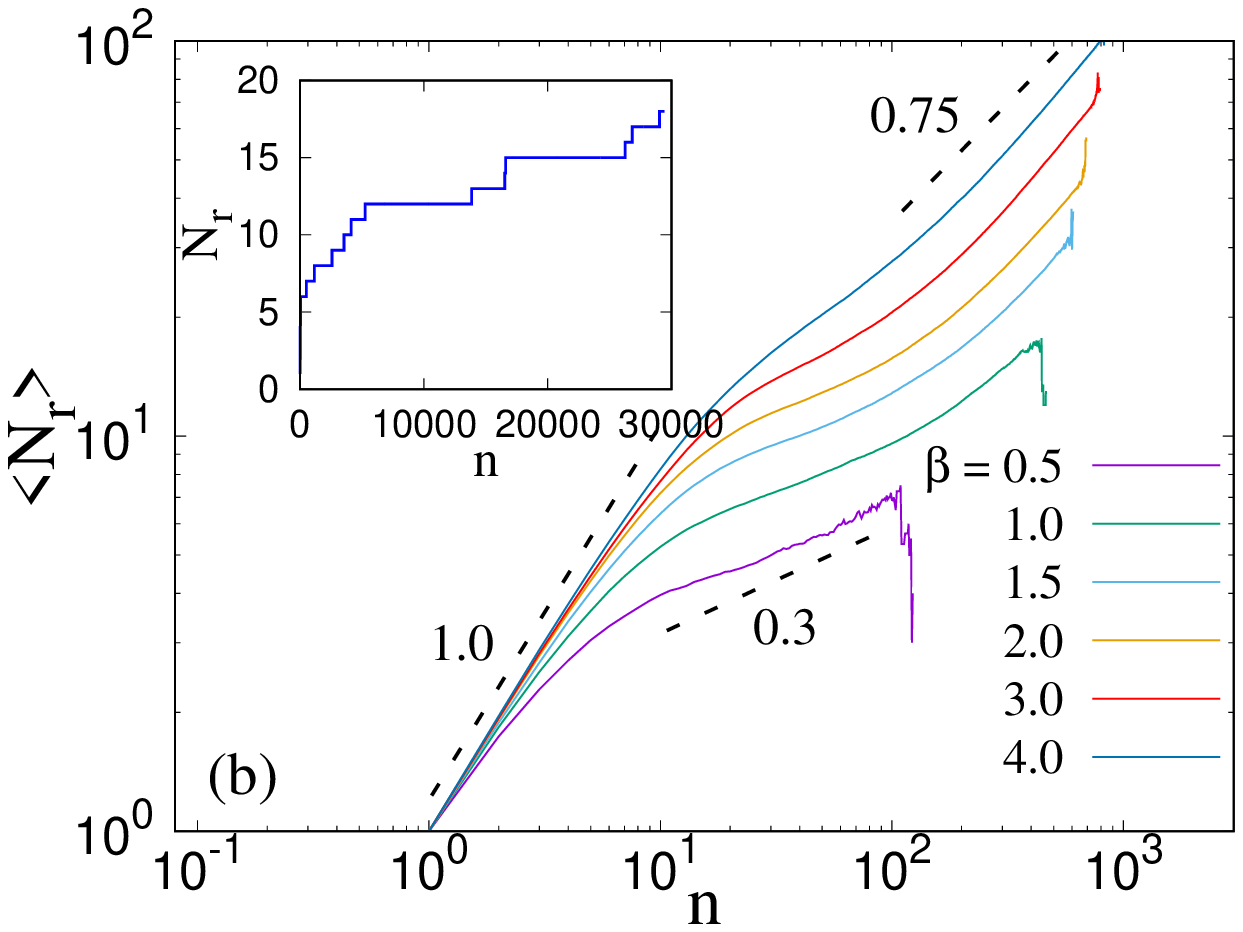}
\caption{(a) Energy emitted during the failure process vs the number of steady states. The red dots correspond to the record events with magnitude $r_k$ corresponding to the $k$th event. $\tau_k$ is the waiting time to achieve the $k$th record event. The figure shows $r_k$ for $k=10$ and $\tau_k$ for $k=14$. (b) Variation of the average number $\langle N_r \rangle$ of record events for a certain number $n$ of total avalanches for $\beta$ ranging between 0.5 and 4.0. Initially, $\langle N_r \rangle \approx n$ as a certain event has a higher probability to be greater than the last one. Later the process slows down and speed-up again closer to the onset of acceleration.} 
\label{fig1}
\end{figure}

We start by observing the nature of the energy emitted as the model evolves with increasing applied stress. Figure \ref{fig1}(a) shows this variation of emitted energy $E$ as a function of the number $n$ of the steady states. Each vertical line on the diagram represents the magnitude of the energy emitted during the avalanche corresponding to that particular steady state. As already mentioned in the model description, a steady state is a point during the evolution of the model where external stress is required to initiate further ruptures. From the definition of an `{\it avalanche}', $n$ can stand for the number of avalanches as well. The size $s$ of an avalanche is the number of fibers broken in between two stress increments (and hence in between two steady states). The corresponding emitted energy will then be 
\begin{align}
E(s) = \displaystyle\sum_{j=1}^{s} \displaystyle\frac{1}{2}h_j^2
\end{align} 
where $j$ runs from $1$ to $s$ (size of the avalanche) and $h_j$'s are the threshold values of the fibers broken during that avalanche. The term $1/2h^2_j$ is the damage energy for the rupture of the jth fiber. A recent article \cite{rb21} shows the lack of correlation between $s$ and $E$ in presence of local stress concentration in FBM, suggesting that a high energy can be achieved through a small burst and vice versa. We have chosen energy over avalanche for our present study since it is a quantity that can be captured easily during an acoustic emission (AE) experiment. The red dots in the same figure denotes the record events: an event that is larger than the last one. The magnitude $k$ of a record event is given by $r_k$ while $\tau_k$ is the waiting time for that event. $\tau_k$ can also be defined as the lifetime of the $(k-1)th$ event. One main aim of the article is to study the behavior of $\tau_k$ and this can be used to predict the failure point. 

Figure \ref{fig1}(b) shows the variation of average number of record events $\langle N_r \rangle$ increases with the number $n$ of the total avalanches. Initially, $\langle N_r \rangle$ increases linearly with $n$. This happens as almost every avalanche emits record energy due to the fact that the reference for record events is very low at the beginning. Later $\langle N_r \rangle$ slows down and increases again close to the failure point with a faster rate but slower than the initial linear increment. This final increased rate of $\langle N_r \rangle$ suggests the onset of acceleration when the model is closer to global failure. Inset of figure \ref{fig1}(b) shows the same behavior but for a single configuration. The vertical jumps take place whenever there is a new record event and the horizontal lines corresponds to no activity zone and hence contributes to the waiting time. The figure explicitly shows that there are a lot of activities with a number of vertical jumps in the beginning as well as at the end while somewhere in the middle, mostly horizontal lines are observed leading to a peak in the waiting time. Such a slowdown can also be observed during a seismic event, a phenomenon widely known as the {\it seismic quiescence}. The acceleration after the quiescence and just before the main shock has been measured quantitatively in the literature by the decrease in slope in the distribution of fore-shock magnitudes and famously known as the {\it b-value} decrease \cite{wchhck18}. Such sequential record events can be characterized as IID sequences which universally shows a logarithmic increment \cite{w13}. Such an increment has also been observed in the context of fiber bundle model recently \cite{kpk20}. In this present article a scale-free increase is observed rather a logarithmic one, most probably due to the lack of correlation between avalanches and energies in the presence of local stress concentration \cite{rb21,bbr22}.    
 
\begin{figure}[ht]
\centering
\includegraphics[width=7.3cm, keepaspectratio]{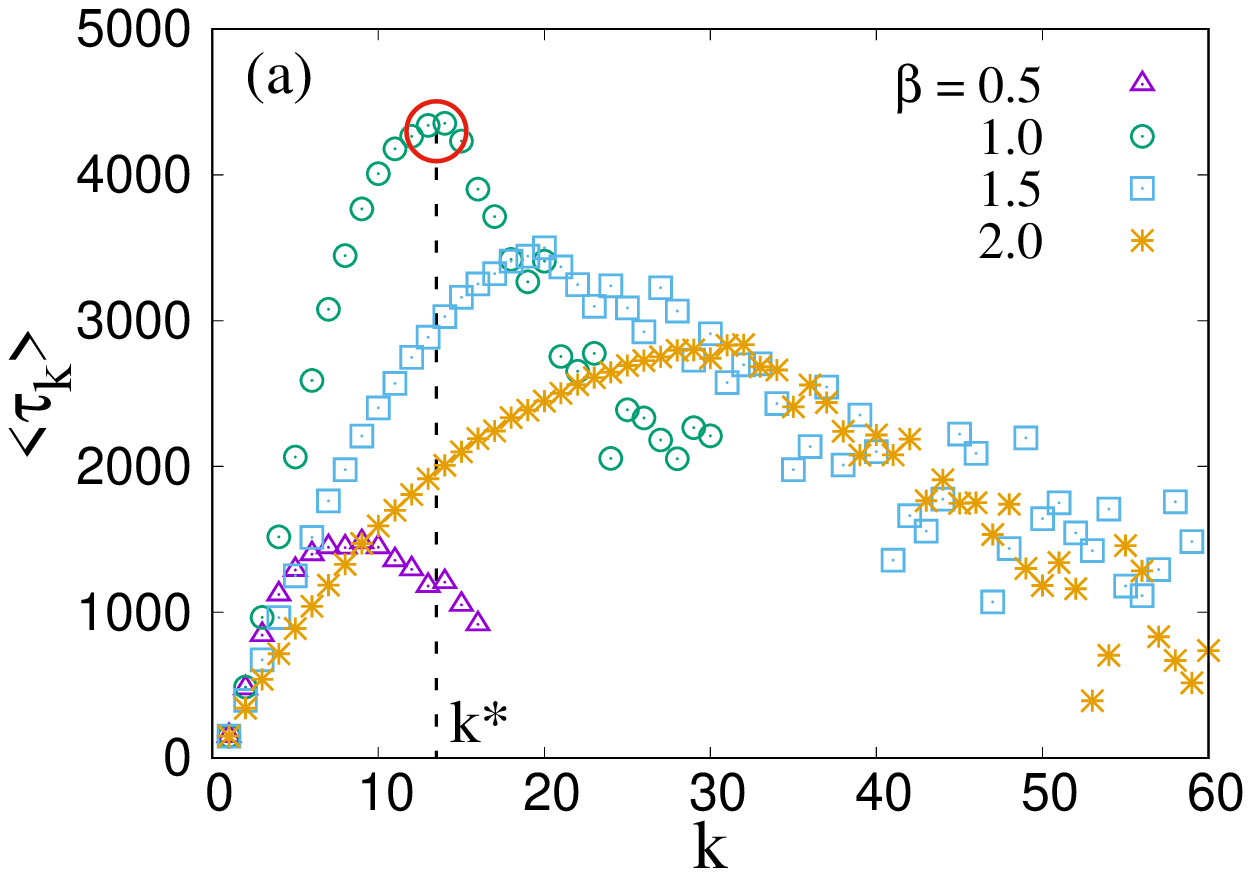} \ \ \ \   \hspace{0.3cm} \includegraphics[width=7.1cm, keepaspectratio]{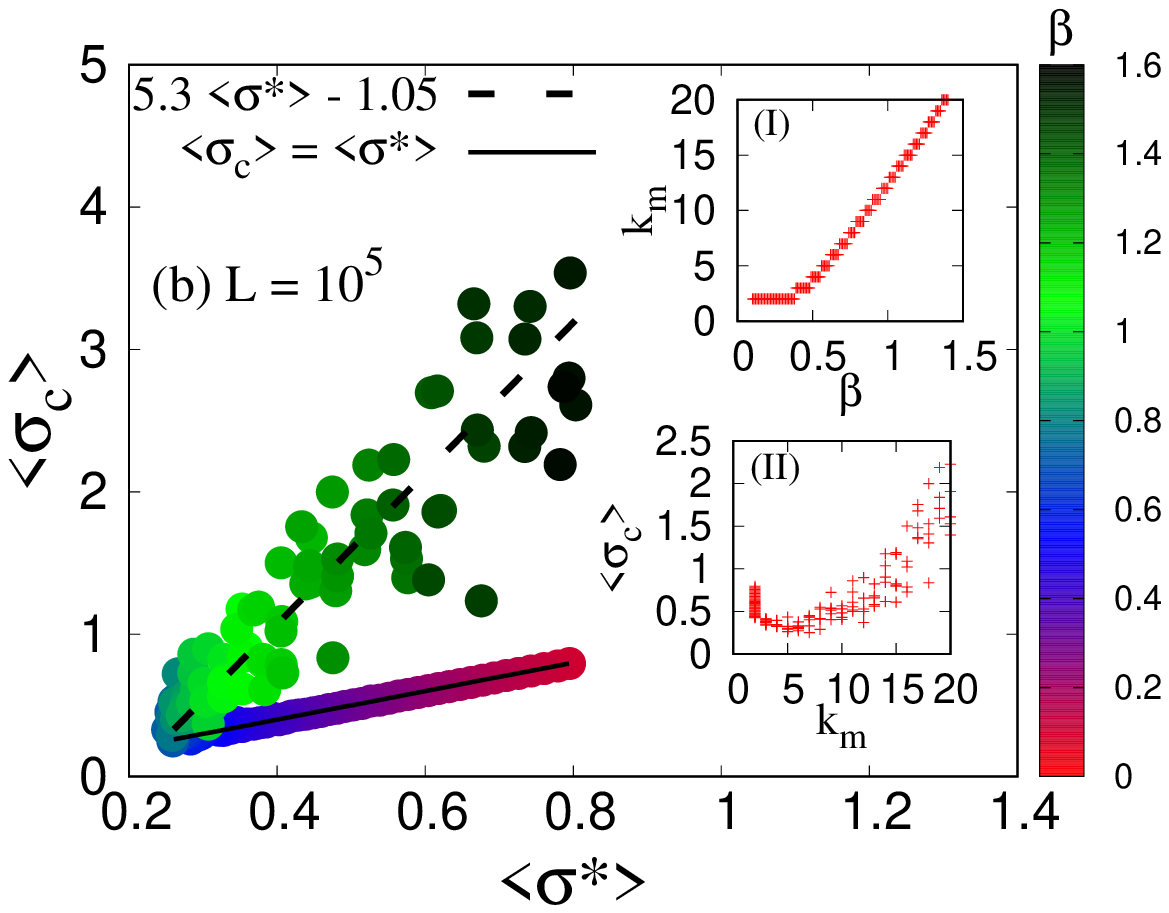}
\caption{(a) Average (over $10^4$ realizations) waiting time $\langle \tau_k \rangle$ as a function of the order $k$ of the record event. $\langle \tau_k \rangle$ increases initially with $k$, and decreases again prior to the failure point. $\langle \tau_k \rangle$ reaches its maximum at $k=k^{\ast}$. (b) Average critical stress $\langle \sigma_c \rangle$ plotted against $\langle \sigma^{\ast} \rangle (=\langle \sigma \rangle \ \text{at} \ k=k^{\ast})$ for $L=10^5$. The color gradient shows the values of $\beta$. At low $\beta$, $\langle \sigma_c \rangle = \langle \sigma^{\ast} \rangle$, suggesting no scope of prediction. For moderate $\beta$, a prediction can be done since we observe $\langle \sigma_c \rangle=a\langle \sigma^{\ast} \rangle - b$, where $a=5.3 \pm 0.18$, and $b=1.05 \pm 0.81$. Inset (I) and (II) shows how $k_m$, the maximum number of record events registered, is related to disorder strength $\beta$ and critical stress $\sigma_c$ respectively.} 
\label{fig3}
\end{figure}

Next, we turn to study the average waiting time $\langle \tau_k \rangle$ as a function of the index $k$ of record events. The averaging is done for each $k$ value over $10^4$ realizations. Figure \ref{fig3}(a) shows $\langle \tau_k \rangle$ as a function of $k$ for $\beta=0.5$, 1.0, 1.5 and 2.0. $\langle \tau_k \rangle$ shows a non-monotonic behavior with a peak at a certain value of $k$, say $k^{\ast}$. On both sides of $k^{\ast}$, $\tau_k$ decreases suggesting that we will see closely spaced record events as we move away from $k^{\ast}$. For $k<k^{\ast}$, this happens since the reference frame that decides an event to be the record one is itself small. On the other side ($k>k^{\ast}$), though the reference is quite large, a decrease in $\tau_k$ suggests we get many bigger events within that limit. This $k^{\ast}$, as we will see later in this paper, can be connected to the failure point predicting an upcoming catastrophic event. 
\begin{figure}[ht]
\centering
\hspace{-0.8cm} \includegraphics[width=7.2cm, keepaspectratio]{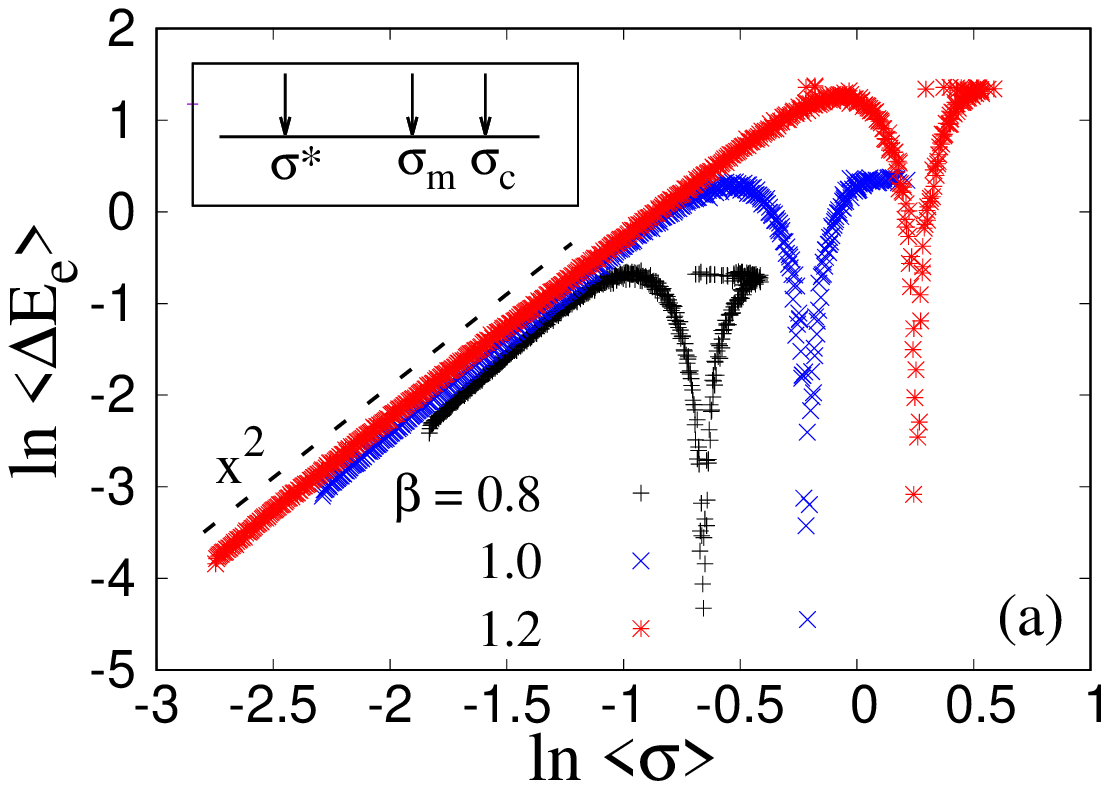} \ \ \ \   \hspace{-0.2cm} \includegraphics[width=7.2cm, keepaspectratio]{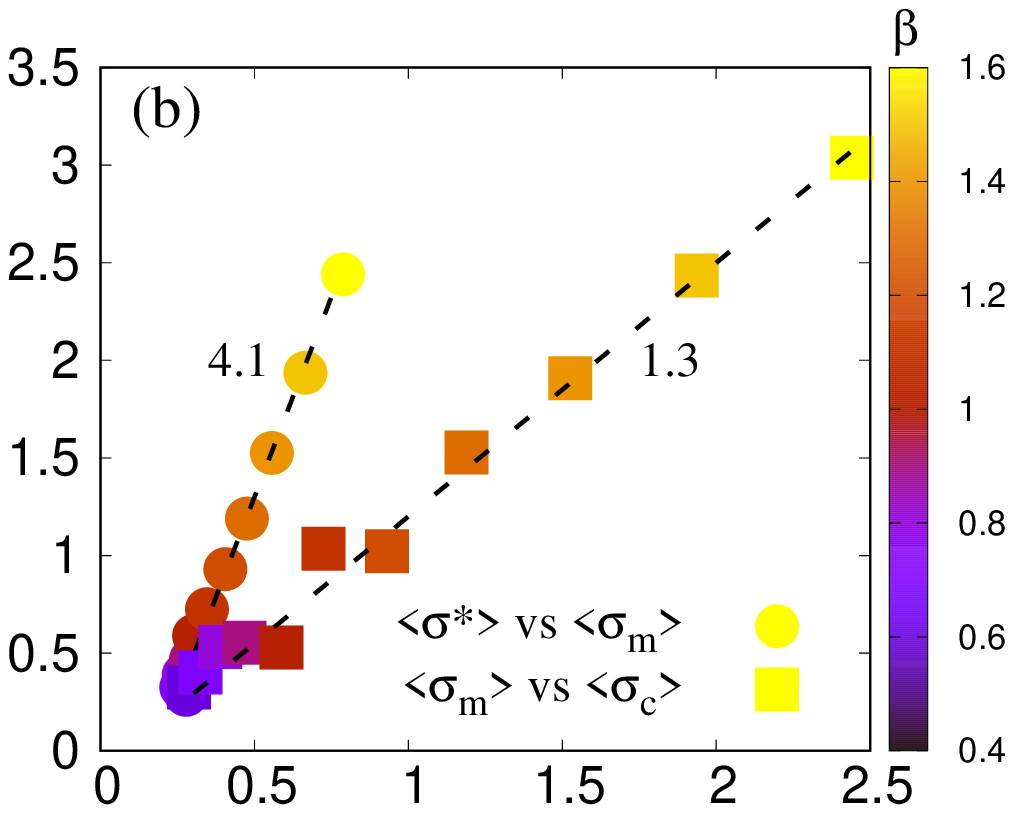}
\caption{(a) Change in average elastic energy $\Delta E_e$ as a function of the external stress for $\beta=0.8$, 1.0 and 1.2. $\Delta E_e$ shows a non-monotonic behavior with a peak at $\sigma_m$. For a certain $\beta$, such peak takes place before $\langle \sigma_c \rangle$ and after $\langle \sigma^{\ast} \rangle$, as it is shown in the inset. (b) The figure shows how $\langle \sigma_m \rangle$ varies with $\langle \sigma^{\ast} \rangle$ and $\langle \sigma_c \rangle$ as $\beta$ is continuously varied. We observe: $\langle \sigma_m \rangle = p\langle \sigma^{\ast} \rangle - q$ and $\langle \sigma_c \rangle = c\langle \sigma_m \rangle - d$, where $p=4.1 \pm 0.09$, $q=0.75 \pm 0.04$, $c=1.3 \pm 0.04$, and $d=0.1 \pm 0.05$.} 
\label{fig4}
\end{figure}
Figure \ref{fig3}(b) shows the variation in $\langle \sigma_c \rangle$ and $\langle \sigma^{\ast} \rangle$, where $\langle\rangle$ denotes an average over $10^4$ realizations. $\langle \sigma_c \rangle$ is the average critical stress which is applied externally on the bundle just before the global failure. $\langle \sigma^{\ast} \rangle$ is the stress at which $\tau_k$ reaches its maximum value. So, $\langle \sigma^{\ast} \rangle = \langle \sigma \rangle_{k=k^{\ast}}$.         

We study the correlation between $\langle \sigma^{\ast} \rangle$ and $\langle \sigma_c \rangle$. Each point, within the figure, is for a certain value of $\beta$ ranging between 0 and 1.6 (see the color gradient). We see two branches of data. The first branch corresponds to low disorder strength when the failure process is brittle-like abrupt \cite{r17,r21}. Here, $\langle \sigma^{\ast} \rangle = \langle \sigma_c \rangle$, which means the peak in $\langle \tau_k \rangle$ takes place at the same time the model reaches the global failure. A prediction will not be possible in this case. However, for high disorder, we observe $\langle \sigma_c \rangle$ to increase linearly with $\langle \sigma^{\ast} \rangle$: 
\begin{align}
\langle \sigma_c \rangle = a\langle \sigma^{\ast} \rangle - b
\end{align}  
where $a=5.3 \pm 0.18$, and $b=1.05 \pm 0.81$. This means, $\langle \sigma_c \rangle$ has a value much higher than $\langle \sigma^{\ast} \rangle$ and we can have information about the failure point much before the failure occurs. The inset of figure \ref{fig3}(b) shows how $k_m$, the maximum number of record events registered varies with disorder strength $\beta$ and the critical stress $\sigma_c$ respectively. A higher $k_m$ suggests the bundle is relatively stable as the final acceleration starts at a higher time. This agrees with how $k_m$ behaves with $\sigma_c$ - a higher critical stress when more record events are captured, supporting the higher stability at higher disorder strength.     

We also want to take this opportunity to explore a different technique to predict the critical stress which was suggested recently \cite{pkh19} (but in the context of the mean-field limit) and see whether it is consistent with our hypothesis. For this, we will study the elastic energy associated with system evolution. The elastic energy at a certain applied stress is defined as:
\begin{align}
E_e = \displaystyle\sum_{j} \displaystyle\frac{1}{2}[\sigma(j)]^2
\end{align}  
where $j$ runs over all intact fibers and $\sigma(j)$ is the local stress of fiber $j$. $E_e$ will then be a non-monotonic function since there are many fibers in the beginning that are less strained while at the end there is a very less number of fibers that are highly strained. Unfortunately, such a peak in $E_e$ takes place after $\langle \sigma_c \rangle$ when the model has already become unstable \cite{pkh19}. Instead, if we calculate the change in elastic energy $\Delta E_e$ as we go from one steady state to the next one, we observe a maximum value of $\Delta E_e$ at $\langle \sigma_m \rangle$, which takes place before $\langle \sigma_c \rangle$. 
\begin{figure}[ht]
\centering
\includegraphics[width=7.5cm, keepaspectratio]{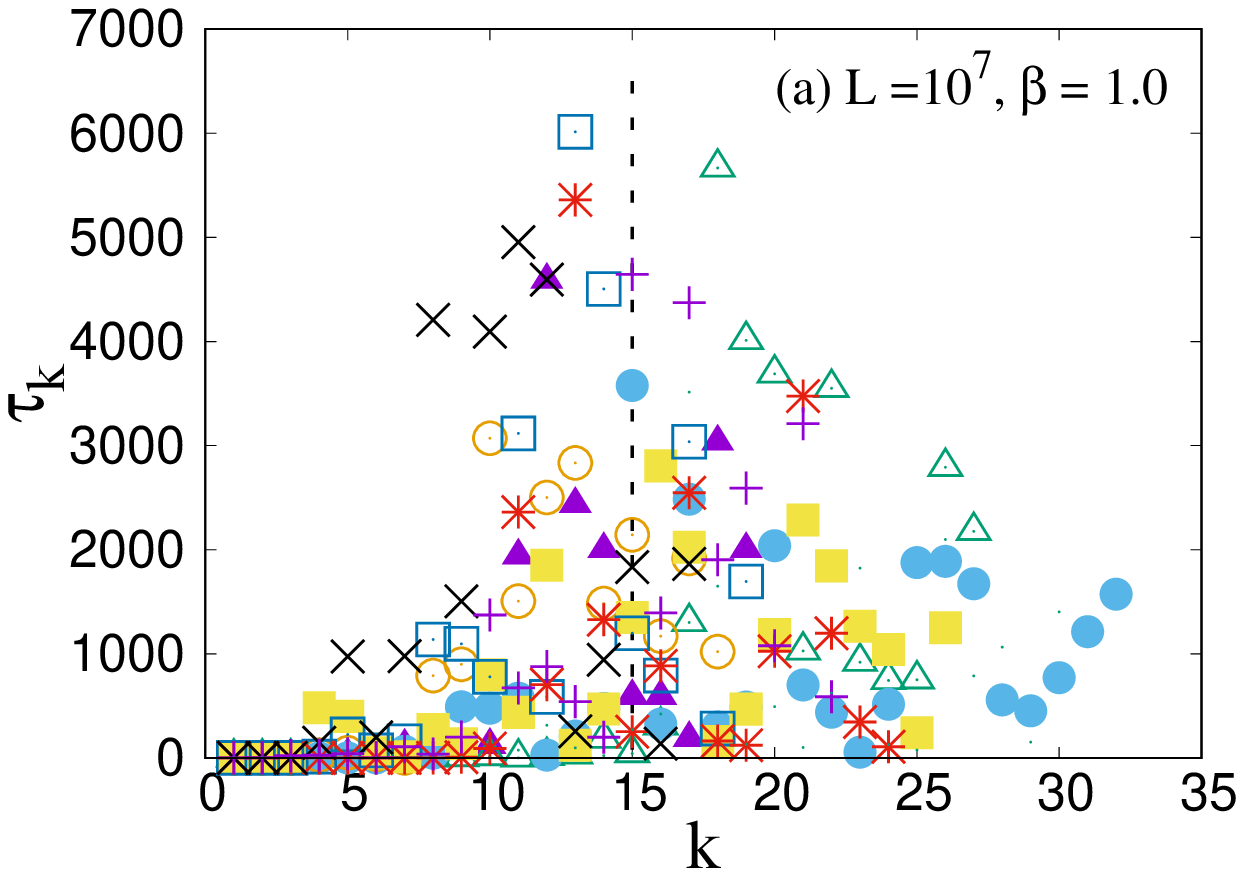} \ \ \ \   \includegraphics[width=7.0cm, keepaspectratio]{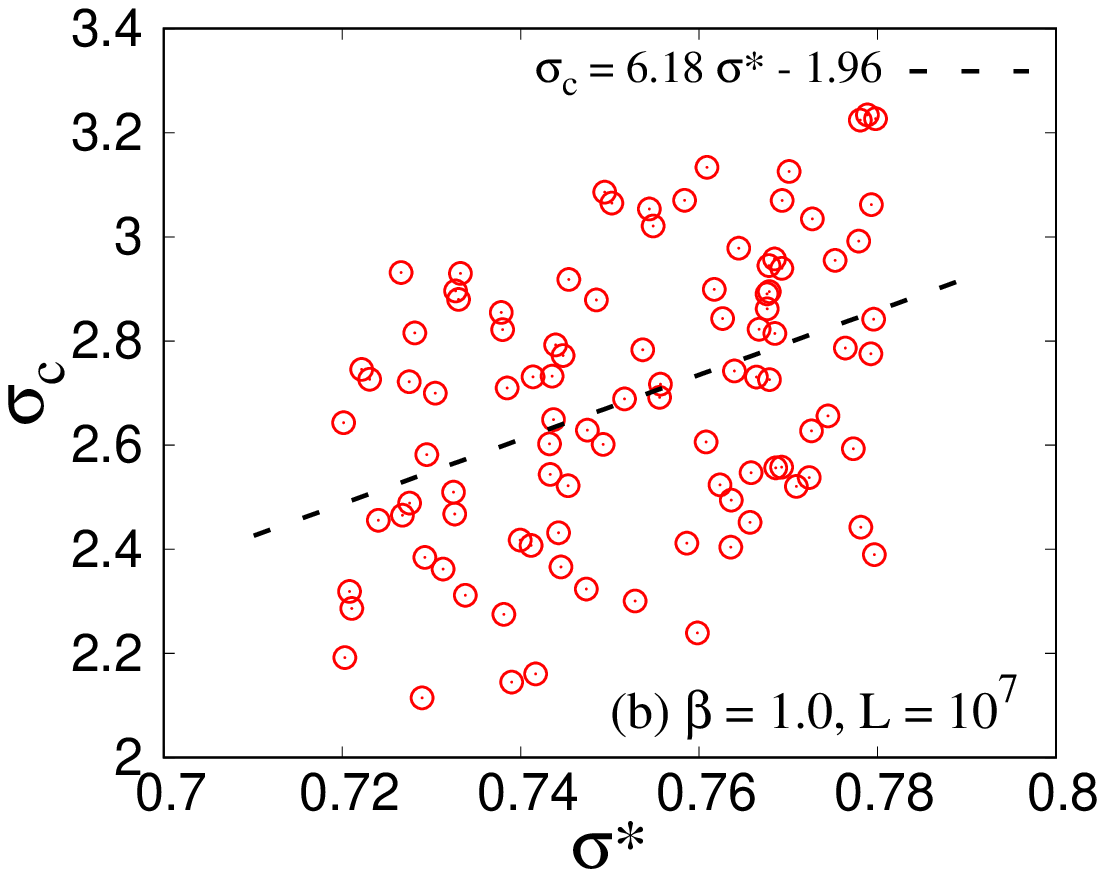} 
\caption{(a) $\tau_k$ as a function of $k$ for a 10 different realizations for a bundle of size $10^7$. $\tau_k$ shows a peak in all the cases making our hypothesis valid not only for average parameters but for a single realization as well. (b) Prediction of $\sigma_c$ from $\sigma^{\ast}$ for 100 different configurations for $\beta=1.0$. We observe: $\sigma_c=a^{\prime}\sigma^{\ast}-b^{\prime}$  from least square fit, where $a^{\prime}=6.18 \pm 1.37$ and $b^{\prime}=1.96 \pm 0.94$.} 
\label{fig5}
\end{figure}

Figure \ref{fig4}(a) shows $\Delta E_e$ to increase as $\langle \sigma \rangle^2$ prior to the peak independent of $\beta$ and decreases sharply afterward. The inset of the same figure shows schematically where $\langle \sigma^{\ast} \rangle$, $\langle \sigma_m \rangle$ and $\langle \sigma_c \rangle$ lie. The difference between $\langle \sigma_m \rangle$ and $\langle \sigma_c \rangle$ is very less while $\langle \sigma^{\ast} \rangle$ takes place much early (see the inset). So, $\langle \sigma^{\ast} \rangle$ serves as a better prediction point rather than $\langle \sigma_m \rangle$ but we can treat it to be the second warning point prior to global failure. Figure \ref{fig4}(b) shows the following correlation
\begin{align}
\langle \sigma_m \rangle =  p\langle \sigma^{\ast} \rangle - q \ \ \text{and} \ \
\langle \sigma_c \rangle =  c\langle \sigma_m \rangle - d
\end{align}  
where where $p=4.1 \pm 0.09$, $q=0.75 \pm 0.04$, $c=1.3 \pm 0.04$, and $d=0.1 \pm 0.05$. Above correlation shows that not only $\langle \sigma_c \rangle$ but $\langle \sigma_m \rangle$ also can be predicted if we know $\langle \sigma^{\ast} \rangle$.

Finally, we will discuss our hypothesis in the context of a practical application. So far, we have only discussed the average quantities which are not useful when comparing with experiments since a single experiment is equivalent to a single realization. Here, we chose a $\beta$ value large enough ($=1.0$) so that the prediction is possible. Figure \ref{fig5}(a) shows the variation in $\tau_k$ with index $k$ for 10 different realizations for $L=10^7$ and $\beta=1.0$. For each realization, $\tau_k$ consistently shows a peak before the global failure and we can extract both $\sigma^{\ast}$ and $\sigma_c$ for each realization. Figure \ref{fig5}(b) shows how $\sigma_c$ varies with $\sigma^{\ast}$ for 100 such realizations. The dotted line shows the best fit according to the least square method:
\begin{align}
\sigma_c = a^{\prime}\sigma^{\ast} - b^{\prime}
\end{align}  
where where $a^{\prime}=6.18 \pm 1.37$ and $b^{\prime}=1.96 \pm 0.94$. The above correlation shows that the prediction of critical stress $\sigma_c$ is possible for individual configurations as well.


\section*{Discussion}
When a disordered system is acted upon a constant stress (creep \cite{rh18,rh20}) or a slowly varying external load, the process leads to a jerky evolution of the fracture process which is often detected externally by a sequence of acoustic signals. The present work is mainly based on such a sequence of acoustic signals where the record events (most prominent ones) are the center of interest. The paper explores an efficient method through studying the statistics of the record signals to predict the onset of an accelerating nature and eventual catastrophic failure.   

As an important outcome of the work, I have represented that during quasi-brittle fracture, by combining the record statistics with avalanche size distribution during a failure process it is possible to predict the failure point with high accuracy. The technique is built on the behavior of waiting times between consecutive record bursts. Such waiting-time shows a non-monotonic behavior where the value of applied stress at which the waiting time is maximum shows a high correlation with critical stress. Such a technique has a two-fold purpose: first, a warning of an upcoming failure through the maximum value of waiting time as well as the maximum change in elastic energy, and second, correlating the critical point with the stress/strain at this maxima. The average applied stress $\langle \sigma^{\ast} \rangle$ at maximum waiting time, as well as the average applied stress $\langle \sigma_m \rangle$ at the maximum change in elastic energy, are observed to be correlated linearly with the average critical stress $\langle \sigma_c \rangle$. One challenge is such an early warning of imminent failure requires to be done much before the model enters the unstable state. Since $\langle \sigma^{\ast} \rangle$ takes place much before $\langle \sigma_m \rangle$, the process with the record statistics gives us a larger time for damage control than the elastic energy treatment. As it was already observed in the literature, a disordered system like the fiber bundle model shows brittle-like abrupt failure \cite{rr15,r17,r21,rbr17} (in a single avalanche) when the disorder strength is low. Clearly, this hypothesis fails in this limit of low disorder, where the failure process is brittle-like abrupt \cite{rr15,r17,r21,rbr17}, due to a lack of avalanche statistics. Particularly, it was shown that at low disorder $\langle \sigma^{\ast} \rangle$ is almost equal to $\langle \sigma_c \rangle$, and a prediction is not possible. That's why this treatment is only fruitful in the quasi-brittle region only. 

This paper is a step closer toward failure prediction from a statistical point of view. It is imperative to extend the study to realistic systems and engineering materials towards understanding the validity of this novel process starting from laboratory scales like glassy materials, and damage mechanics in continuum media to geological scales like Heines jumps during multi-phase flows in water and oil industries as well as large scale seismic events.  


\section*{Methods}
The methods used for the work are numerical. Codes written by the author in C-language are used for the simulations. The fibers are broken following the {\it weakest-link-of-the-chain} theory. If $h_i$'s are the threshold values of the fibers and $\sigma_i$ are corresponding local stress values, then the fiber that breaks has the minimum value of $(h_i-\sigma_i)$. The stress of the broken fiber is redistributed among the neighboring fibers. In one dimension, which I have used for this study, the two neighbors (right and left) have the following local stress: $\sigma_{r(l)} \rightarrow \sigma_{r(l)}+\displaystyle\frac{\sigma_b d_{l(r)}}{d_r+d_l}$, where $\sigma_b$ is the stress of the broken fiber, $d_l$ and $d_r$ are the distance of the neighboring fiber from the broken one on the left and right sides respectively. After the redistribution, some other fibers may break due to the elevated local stress profile starting an avalanche. After an avalanche, the next {\it weakest} fiber is broken by calculating the minimum of $(h_i-\sigma_i)$ again. For sufficiently nice statistics, I have simulated a bundle of size ranging between $10^3$ and $10^7$ with $10^4$ realization.   


\section*{Acknowledgements}
SR acknowledges Birla Institute of Technology \& Science, Hyderabad campus for the support during this work. A special thank goes to Prof. Rajesh Ravindran, Prof. Anuradha Banerjee, Prof. Purusattam Ray, and Subrat Senapati for discussions and valuable comments from time to time.


\section*{Author contributions statement}
The author confirms being the sole contributor of this work and has approved it for publication. 


\section*{Accession codes} 
The codes can be accessed through the link \href{https://github.com/sroy2807/failurePredictionFBM.git}{https://github.com/sroy2807/failurePredictionFBM.git}.


\section*{Competing interests} 
The author declares no competing financial interests. 



\end{document}